# The Energy Efficiency of Interfacial Solar Desalination: Insights from Detailed Theoretical Analysis

Xiao Luo[1,5], Jincheng Shi[2,5], Changying Zhao[1], Zhouyang Luo[2,3,4,*], Xiaokun Gu[1,*], Hua Bao[2,*]

[1] Institute of Engineering Thermophysics, School of Mechanical Engineering, Shanghai Jiao Tong University, Shanghai 200240, China

[2] University of Michigan-Shanghai Jiao Tong University Joint Institute, Shanghai Jiao Tong University, Shanghai 200240, China

[3] Key Laboratory of Solar Energy Utilization & Energy Saving Technology of Zhejiang Province, Hangzhou 311121, China

[4] Zhejiang Energy R&D Institute Co., Ltd, Hangzhou 311121, China

[5] These authors contributed equally: X. Luo, J. Shi.

* Corresponding authors.

E-mail address: Zhouyang_Luo@outlook.com (Z. Luo); xiaokun.gu@sjtu.edu.cn (X. Gu); hua.bao@sjtu.edu.cn (H. Bao).

**Summary:** Solar-thermal evaporation, a traditional steam generation method for solar desalination, has received numerous attentions in recent years due to the significant increase in efficiency by adopting interfacial evaporation. While most of the previous studies focus on improving the evaporation efficiency by materials innovation and system design, the underlying mechanisms of its energy efficiency are less explored, leading to many confusions and misunderstandings. Herein, we clarify these mechanisms with a detailed thermal analysis model. Using this model, we elucidate the advantages of interfacial evaporation over the traditional evaporation method. Furthermore, we clarify the role of tuning the solar flux and surface area on the evaporation efficiency. Moreover, we quantitatively prove that the influence of environmental conditions on evaporation efficiency could not be eliminated by subtracting the dark evaporation rate from evaporation rate under solar. We also find that interfacial evaporation in a solar still does not have the high overall solar desalination efficiency as expected, but further improvement is possible from the system design part. Our analysis gains insights to the thermal processes involved in interfacial solar evaporation and offers perspectives to the further development of interfacial solar desalination technology.

# 1. Introduction

Solar-thermal evaporation, in which renewable solar energy is utilized to drive thermal evaporation, has been widely used in seawater desalination since ancient time[1,2]. It has been also applied to different industrial applications, such as wastewater treatment[3], power generation[4], and steam sterilization[5]. In traditional solar-thermal evaporation system, such as solar stills[6,7], the solar absorber is placed at the bottom of the bulk water (Fig. 1a), namely bottom heating, which has an evaporation efficiency of 30%-45% due to the separation of heat and vapor generation[8]. To reduce heat loss, volumetric heating using ink[9] or nanofluids suspended in water[10] has been put forward to achieve heating uniformly (Fig. 1b). However, the evaporation efficiency improvement is relatively low and the long-term stability of nanofluids remains challenging[11,12]. In 2014, the concept of interfacial solar evaporation has been introduced successfully to augment the evaporation efficiency to 64% by heat localization at the evaporation interface (Fig. 1c), which has greatly promoted the development of this field due to its high evaporation efficiency[13]. Such an idea has been conceived in earlier applications. For example, the U.S. military used to make a simple first-aid kit with a black wet towel and a plastic bag to generate fresh water for the pilots and sailors who have fallen into the sea during World War II[14], whose mechanism behind could be regarded as interfacial evaporation.

Starting from the first demonstration of interfacial solar evaporation[13], there are extensive follow-up researches[8,15-20] aiming at material innovation and thermal management to boost the evaporation efficiency, which is commonly used to assess the evaporation performance. Nowadays, more than 80% interfacial evaporation efficiency has been routinely achieved[21-31]. Evaporation efficiency more than 100% has also been observed[32-36] by several groups and up to 10.9 kg/(m$^2$·h) under one sun[36] (theoretical upper limit 1.5 kg/(m$^2$·h)) has been recently reported[32]. On the other hand, although desalination is regarded as the most promising application of interfacial solar evaporation, there are limited researches focusing on fresh water harvesting rate[37-39]. Especially, when the interfacial evaporation system is placed within a solar still, the performance seems to be lower than expectation[29,31,40-42].

Despite the rapid evolution of interfacial solar evaporation research, several questions arise from reviewing different experimental results and viewpoints.

First, it seems intuitive to believe that interfacial heating has better evaporation performance than the more traditional volumetric heating and bottom heating technology since the heat is localized. However, for volumetric and bottom heating, the solar energy is not localized on the surface but still fully absorbed by the bulk water, which should not be regarded as a heat loss, because the thermal stored in the bulk water would eventually be released by evaporation. This is also pointed out by Pang *et al.* in a recent review[15]. Therefore, what are the fundamental advantages of the interfacial evaporation in comparison to the traditional methods?

Second, many studies indicated that low evaporation temperature by enlarging the evaporation surface area (equivalently, reducing the solar flux per unit area) can enhance the efficiency due to the less thermal losses[32-35,43,44]. Some researchers further developed structures to gain environmental energy by lowering the evaporation temperature to below the environmental temperature, through which the evaporation rate can even excess the theoretical upper limit[32-35,43,44]. However, some other researches showed that high evaporation temperature by concentrating solar energy would contribute to high evaporation efficiency[13,15]. How can we understand the seemingly conflicting results from these studies?

Third, to compare the evaporation performances of different structures by different groups, most previous studies aim at eliminating the effect of environmental conditions by generally subtracting the dark evaporation (i.e., the evaporation without solar) from the total evaporation under solar to obtain the "pure solar-induced evaporation"[13,45]. However, several studies indicate that the dark evaporation should not a part of the evaporation under solar[34,46]. Hence, a natural question is how we can fairly evaluate the evaporation efficiency under different environmental conditions?

Fourth, in the open space, the evaporation efficiency has been reported to be as high as about 80% and the evaporation rate can reach 1.2 kg/(m$^2 \cdot$h)[21-23]. However, several studies showed that the water production efficiency of an actual solar still, which is an enclosed space with interfacial evaporation structures is only about 40%[29,31,40-42,47], which is similar to the traditional bottom heating solar still[48-50]. What is the mechanism of heat loss and how could one further improve the water production efficiency of solar still based on interfacial evaporation structure?

The above questions indicate that better understanding of the underlying mechanisms of interfacial solar evaporation are still highly desirable.

In this work, we address the above questions through a detailed theoretical analysis, which can provide insights into the fundamental mechanisms of interfacial solar evaporation. We set up a model based on energy balance equation and heat/mass transfer models to accurately simulate the energy transfer in the interfacial solar evaporation system, as well as a single stage solar desalination system. By theoretical analysis, we identify the advantages of interfacial heating over volumetric heating and the determining factors for the interfacial evaporation efficiency. Furthermore, we quantitatively illustrate the effect of the environmental conditions to the interfacial evaporation efficiency, so that one can fairly evaluate the evaporation performance under different environmental conditions. In addition, interfacial evaporation in solar still is modeled and the energy loss mechanisms are clarified.

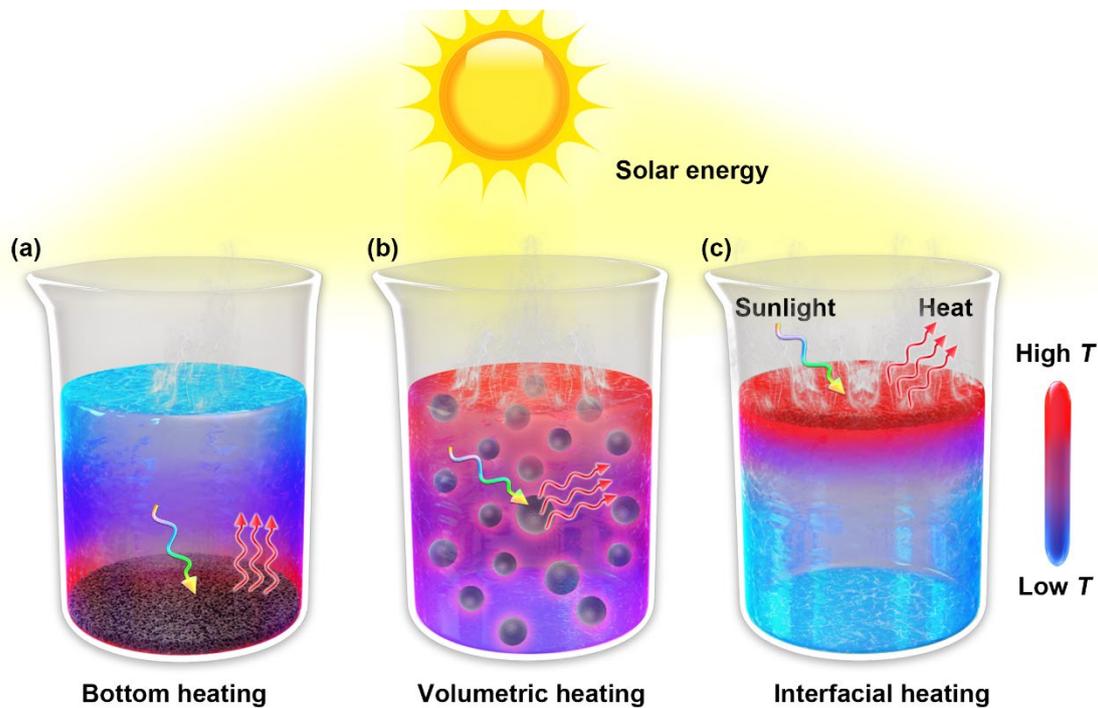

**Fig. 1. Different types of solar absorbers studied in the solar-thermal water evaporation field.**
(a) Bottom heating; (b) Volumetric heating; (c) Interfacial heating.

## 2. Models

The typical interfacial solar evaporation system consists of two functional components, as shown in Fig. 2a. The top evaporation layer is the part to absorb solar irradiation and convert it into thermal energy, while allowing the vapor to escape from the surface. Beneath the solar absorber is a thermal insulating component with water channels, which can simultaneously avoid downward heat dissipation and allow water flow to the evaporation layer by capillary force. The two functional components can be realized by two or more separated materials[13,45,51] or a single piece[30,40,52,53] of material.

Since volumetric heating would have a better performance than the bottom heating due to the uniform heating[54-56], here we only develop theoretical models for interfacial heating and volumetric heating (assuming uniform heat generation) to understand the fundamental advantages of the interfacial evaporation in comparison to the traditional methods, as Fig. 2b shows. The evaporation performance of interfacial heating is mostly independent of the thickness of the underlying bulk water because of the insulation layer beneath the evaporation layer. Thus, we regard the bulk water as a heat sink for interfacial heating in our model, i.e., the bulk water temperature remains at environmental temperature. However, the performance of volumetric heating is related to the bulk water thickness. Thus, it is necessary to choose different thickness for volumetric heating to justify the effect of the bulk water thickness on evaporation efficiency. In this work, the thicknesses are chosen as 1 cm, 5 cm, and 10 cm, as Fig. 2b shows. Bulk water is placed in a thermal insulation container to reduce the side and bottom heat loss, as Fig. 2c and 2d shows.

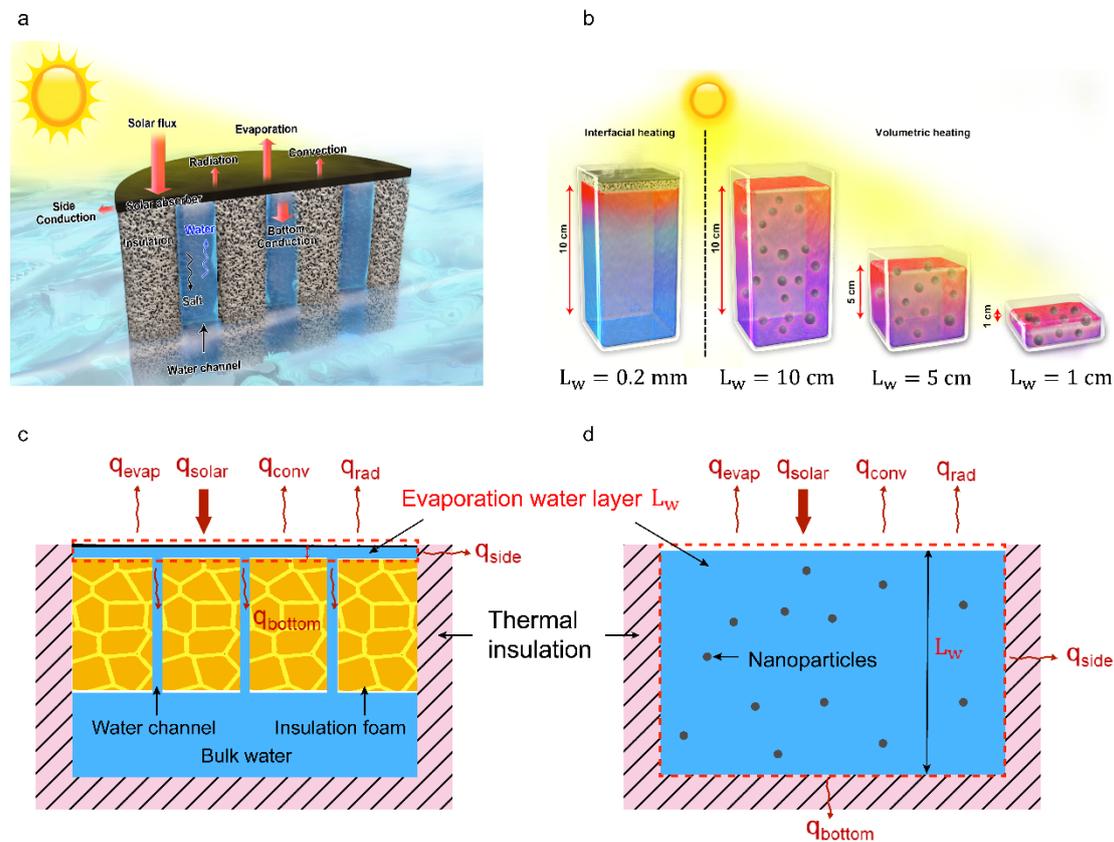

**Fig. 2. Schematic diagram of interfacial solar evaporation system.** (a) Typical interfacial solar evaporation structure. The top layer is a solar absorbing layer. Beneath the solar absorbing layer is a thermal insulation layer. (b) Evaporation structure for interfacial heating and volumetric heating. The evaporation water layer thickness is 0.2 mm for interfacial heating; the evaporation water layer thicknesses are 10 cm, 5 cm, and 1 cm for volumetric heating. (c) Side view of the evaporation structure for interfacial heating. (d) Side view of the evaporation structure for volumetric heating.

Figure 2c and 2d also show all heat transfer processes involved in the evaporation water layer for interfacial heating and volumetric heating, respectively, including solar flux ($q_{solar}$), evaporation heat flux ($q_{evap}$), convection ($q_{conv}$), radiation ($q_{rad}$), and conduction ($q_{cond}$) heat losses. The conduction heat loss ($q_{cond}$) includes the side conduction ($q_{side}$) to the ambient, and bottom conduction ($q_{bottom}$) to the underlying bulk water, i.e., $q_{cond} = q_{side} + q_{bottom}$. All heat fluxes above are normalized by the top projected area[46].

In order to answer the above questions and further explain the heat transfer mechanisms of interfacial solar evaporation, here we develop a detailed heat and mass transfer model. Different from the steady-state analysis in previous works[57,58], we emphasize that the evaporation under natural sunlight is a transient process[47]. The transient energy balance of the evaporation water layer for interfacial heating and volumetric heating can be both expressed as,

$$\rho_w c_p L_w \cdot \frac{dT_{evap}}{dt} = \alpha \cdot q_{solar} - q_{evap} - q_{conv} - q_{rad} - q_{side} - q_{bottom}, \quad (1)$$

where $\rho_w$ is the water density, $c_p$ is the water specific heat, $L_w$ is the evaporation layer thickness, $T_{evap}$ is the evaporation temperature, $t$ is time, $\alpha$ is the solar absorptivity of the absorbing layer. Each term in Eq. (1) is described specifically in the following.

The convection energy $q_{conv}$ can be described as,

$$q_{conv} = h_{conv}(T_{evap} - T_\infty), \quad (2)$$

where $h_{conv}$ is the natural convection heat transfer coefficient between water surface and environment, $T_{evap}$ is the evaporation temperature, and $T_\infty$ is the environmental temperature.

The mass transfer process by evaporation can be analogous to the convection heat tranfer process[59]. Thus, it is important to figure out the relationship between the convection energy ($q_{conv}$) and evaporation energy ($q_{evap}$). The evaporation process is driven by the vapor pressure gradient. The vapor pressure at the evaporation surface is the saturated vapor pressure corresponding to the evaporation temperature $T_{evap}$. The vapor pressure at infinity is the environmental vapor pressure, which is related to the environmental relative humidity $\varphi$ and the environmental temperature $T_\infty$. By the analogy to the convection heat transfer process, there is a general relation for evaporation energy based on Dunkle's model[60]:

$$q_{evap} = 0.01623 \times h_{conv} \left( e^{25.317 - \frac{5144}{T_{evap}}} - \varphi e^{25.317 - \frac{5144}{T_\infty}} \right). \tag{3}$$

The natural convection heat transfer coefficient $h_{conv}$ can be determined by the special Nusselt and Grashof number in Dunkle's model in Ref. [60] (details in ESI S1).

The side conduction heat losses ($q_{side}$) are quite different for interfacial heating and volumetric heating, because the side surface area of the evaporation water is quite different. The side conduction heat losses are calculated in ESI S1 by using the series thermal resistance model. The influence of side conduction losses to the evaporation efficiency for interfacial heating and volumetric heating will be discussed in details later.

As has been mentioned above, for interfacial heating, beneath the evaporation water is an insulation layer with water channel (Figure 2c). However, for volumetric heating, beneath the evaporation water is a wrapped insulation material (Figure 2d). Thus, the bottom conduction heat losses are different for interfacial heating and volumetric heating, which are also calculated by using the parallel and series thermal resistance model, respectively, whose expressions are documented in ESI S1.

The radiation heat transfer between the evaporation surface and its surroundings is complicated for indoors experimental test, which depends on experimental setups and surroundings. It should be pointed out that, in the real indoor experiments, the lens of the solar simulator is directly above the evaporation surface, which has a higher temperature than the evaporation surface. In this case, the radiation loss could be more negligible. Thus, for simplicity, the radiation heat transfer $q_{rad}$ between the evaporation surface and the surroundings is assumed to be $0$[47]. Other assumptions of the parameters in our model are listed in Table S1 in ESI SI, which should be reasonable to describe typical experimental conditions.

## 3. Discussion

## 3.1 Advantages of interfacial evaporation over volumetric evaporation

By numerically solving Eq. (1) for interfacial and volumetric heating model, we can obtain the evaporation energy $q_{evap}$ as a function of time. The evaporation efficiency is defined as $\eta_e = q_{evap}/q_{solar}$. Figure 3a shows the evaporation efficiency variation over time for interfacial heating and volumetric heating under 1 kW/m² irradiation and 0.03 W/(m · K) thermal insulation (commonly used lab conditions). As Fig. 3a shows, the evaporation efficiency increases first (i.e., transient heating period) and then become stable (i.e., steady state). We define the transient heating time as the period from the starting point of heating until the evaporation efficiency change rate smaller than 1%/min. The time period after the transient heating time is thus called the steady-state period. As Fig. 3a shows, the steady-state evaporation efficiency with interfacial heating is about 80%, which is basically consistent with previous studies[45,61] and validates of our model. The transient heating time is extremely short (< 5 min) for interfacial heating, while the transient heating time is much longer for volumetric heating. Furthermore, thicker bulk water would lead to longer transient heating time for volumetric, which is caused by the large heat capacity of the bulk water.

In addition to transient heating time, the steady-state evaporation efficiency is also an important measure of the system. Figure 3b shows the steady-state evaporation efficiency for different structures and different thermal insulation conditions under 1 kW/m² irradiation. When the thermal insulation container is chosen to be a typical thermal insulation material (such as polystyrene foam) with a fixed thickness of 1 cm (details in ESI S1), whose thermal conductivity is 0.03 W/(m·K), the steady-state evaporation efficiency of interfacial heating is much higher than that of volumetric heating (blue bar in Fig. 3b). In addition, thicker bulk water is found to lead to lower steady-state evaporation efficiency, which is attributed to the side conduction loss through the side thermal insulation. In comparison, an adiabatic thermal insulation condition is also considered (orange bar in Fig. 3b). We find that the steady-state evaporation efficiencies are identical for different structures in such an ideal situation. These results indicate that the interfacial solar evaporation does not have advantages if the systems are well insulated. Thus, the advantage of interfacial heating over volumetric heating is that it requires less external thermal insulation.

Apart from the performances in lab conditions where the solar density is fixed at 1 kW/m², it is essential to know their one-day evaporation performances under actual outdoor conditions where the solar density is varied over time. A calibrated solar power meter (CEL-NP2000-2A, Au-Light) is used to continuously measure the outdoor solar flux on a roof of Longbin Building located in the Shanghai Jiao Tong University campus in Shanghai on Aug. 16 2020. Figure 3c shows the fitted curve of the recording solar density using polynomial fitting method, which is used as the input of parameter $q_{solar}$ in Eq. (1). Then, we could get the evaporation energy $q_{evap}$ for different structures with 0.03 W/(m·K) thermal insulation. As Fig. 3c shows, the evaporation energy of interfacial heating increases more quickly and reach the higher evaporation energy than that of volumetric heating. However, when the solar flux decreases, the evaporation energy of interfacial heating also decreases quickly than that of volumetric heating, which is attributed to the less stored heat in the thin evaporation water layer for interfacial heating. Furthermore, when the bulk water with volumetric heating is thicker, the evaporation energy decreases more slowly. The large heat capacity stored in the bulk water would be eventually released by evaporation under the absence of the sunlight. Thus, the heat dissipation to the underlying bulk water is not necessarily a part of heat loss, which is consistent with the viewpoints in Ref. [15].

Figure 3d shows the overall one-day evaporation efficiencies for different structures and different thermal insulation conditions under variable sunlight. An adiabatic thermal insulation condition is also considered for comparison. When the thermal insulation conductivity is 0.03 W/(m·K), the volumetric heating has smaller overall evaporation efficiency than interfacial heating and thicker bulk water would lead to smaller evaporation efficiency. When the thermal insulation is adiabatic, the overall evaporation efficiency is identical for different structures.

From above analysis, we find that the thermal insulation conditions have a great impact on the difference between the interfacial heating and volumetric heating, which is related to the side and bottom losses of the bulk water. To further understand the energy loss, we calculate the energy losses for different structures wrapped with thermal insulation materials with a thermal conductivity of 0.03 W/(m·K), which includes bottom and side conduction losses of the evaporation water layer and the convection loss of the evaporation surface. Figure 3e shows the steady-state energy loss constitutions under 1 kW/m² irradiation. Figure 3f shows the overall one-day energy loss constitutions under variable sunlight. Both figures indicate that the side loss difference is the most significant for different structures, which leads to the difference of evaporation efficiency. The side loss of the bulk water with volumetric heating is larger due to the larger side surface area, thus the larger heat dissipation to the environment through the thermal insulation. Hence, the evaporation efficiency for volumetric heating largely depends on the side thermal insulation conditions.

From the above energy analysis for different evaporation structures, we could conclude that the advantages of interfacial heating over volumetric heating are:

(1) For transient heating period, the transient heating time of interfacial heating is much shorter than the volumetric heating, which could make the evaporation occur rapidly.

(2) For steady-state evaporation period, the steady-state evaporation efficiency of interfacial heating is less sensitive to the thermal insulation conditions, which makes interfacial evaporation more convenient to apply to different conditions.

It should be noted that the evaporation efficiencies are not quite different between volumetric heating and interfacial heating when the side walls are well thermally insulated. Thus, for volumetric heating, it is an effective way to enhance its evaporation efficiency by improving the thermal insulation performance of the side walls or reducing the thickness of the bulk water, to reduce the side heat loss to the environment.

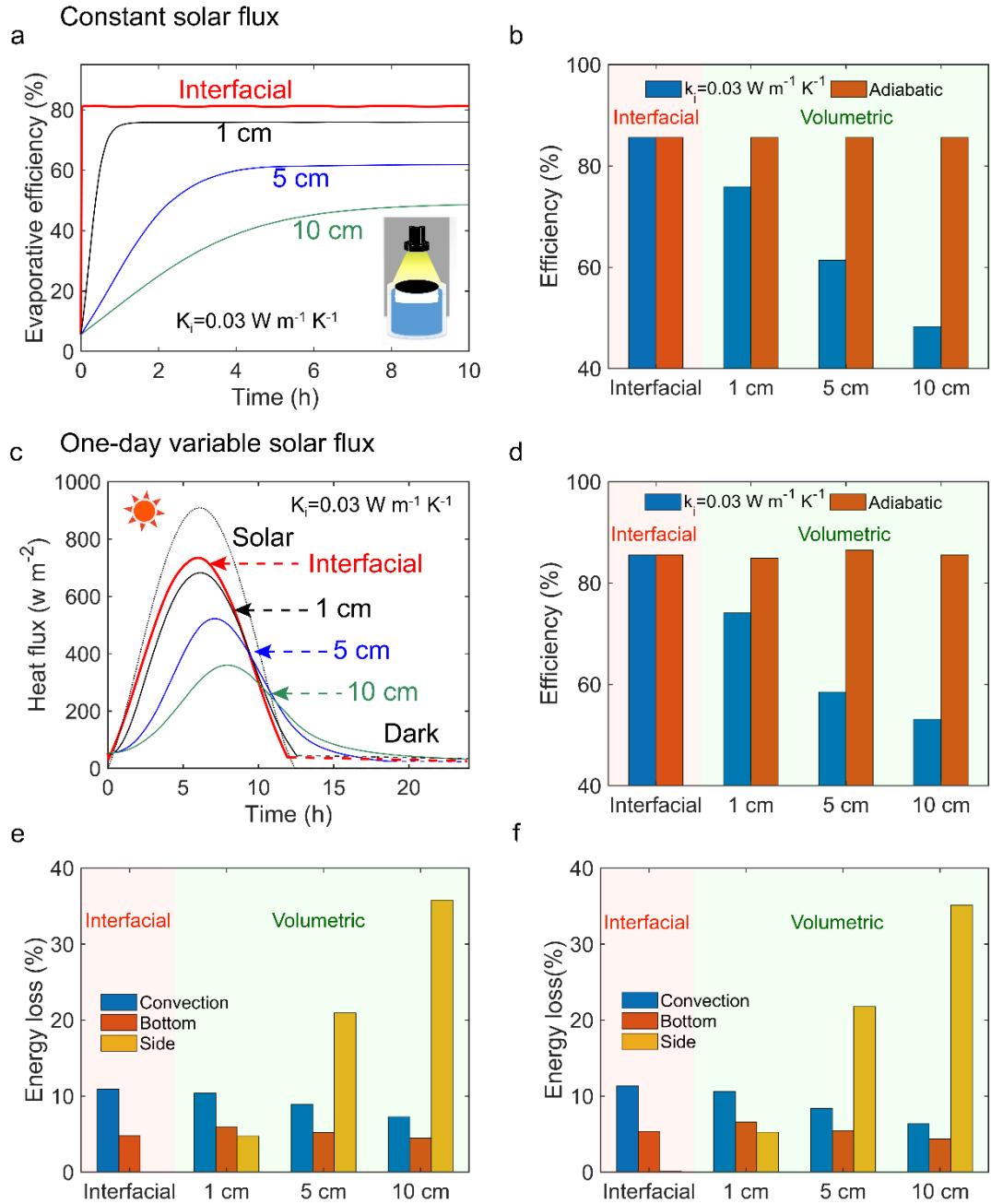

**Figure 3. Performance of interfacial heating and volumetric heating.** (a) Evaporation performance over time under one sun. (b) Steady-state evaporation efficiency under one sun for different thermal insulation conditions (Blue bar is for 0.03 W/(m·K) thermal insulation and orange bar is for adiabatic thermal insulation). (c) Evaporation performance over time under one-day variable sunlight. (d) Overall evaporation efficiency for one-day variable sunlight under different thermal insulation conditions (Blue bar is for 0.03 W/(m·K) thermal insulation and orange bar is for adiabatic thermal insulation). (e) Steady-state energy loss constitutions of interfacial heating and volumetric heating under 1 kW/m$^2$ irradiation. (f) Overall one-day energy loss constitutions of interfacial heating and volumetric heating under one-day variable sunlight.

**3.2 Impact of solar flux to evaporation efficiency**

Some researchers have proved that increasing evaporation surface area (inset in Fig. 4a illustrates one possible way to increase the evaporation surface area) could increase the evaporation efficiency[32,34,43] when the input solar energy is fixed. In essence, increasing the evaporation surface area scatters the fixed solar energy on the increased evaporation surface[32,34], thus decreases the solar flux. Therefore, these studies indicate that decreased solar flux could contribute to higher evaporation efficiency[32,34]. However, on the other hand, some experiments show that increasing solar flux could increase the evaporation efficiency by concentrating solar energy[13,62]. Thus, it is essential to figure out how the solar flux influences the evaporation efficiency.

To extract the relationship between the evaporation efficiency and solar flux, we use variable solar flux as input in Eq. (1). It should be noted that the environmental humidity could also have an influence on the evaporation performance according to Eq. (1) and (3). Thus, to obtain such a relationship applicable to different experimental conditions, we consider the influence of environmental humidity in our calculations, by taking three representative relative humidity 20%, 40%, and 100%. Other assumptions are the same as before, except the solar flux $q_{solar}$.

Figure 4a shows the evaporation efficiency as a function of solar flux under different environmental relative humidity. As shown in Fig. 4a, when $\varphi = 20\%$ and 60%, the evaporation efficiency decreases first and then increases as solar flux increase. Furthermore, the evaporation efficiency could excess 100% due to the energy absorption from the surroundings when the evaporation surface temperature is lower than the environmental temperature. The evaporation efficiency can even be infinite when there is no solar illumination. This is just the convective cooling process, which is a well-known phenomenon[59]. The turning point of the curve is the key to understand the conflicting experimental results in previous studies. Increasing the evaporation surface area could make the average solar flux on the evaporation surface lower than the turning point, thus increasing the evaporation efficiency. This could explain why increasing evaporation surface area could increase the evaporation efficiencies even excessing 100% in previous studies[32,34]. On the other hand, increasing solar flux in previous studies usually focuses on several suns[13,62], which is obviously higher than the turning point, thus eventually increasing the evaporation efficiency. The turning point is also related to the environmental relative humidity. When the humidity increases from 20% to 40%, the turning point decreases. Furthermore, when $\varphi = 100\%$ (saturated humidity), the evaporation efficiency monotonously increases as solar flux increases, which indicates that decreasing solar flux may not necessarily increase the evaporation efficiency in some conditions, such as the solar still ($\varphi = 100\%$).

To better understand the underlying mechanisms, it is essential to further explore the origins that different solar flux could lead to different evaporation efficiency under different environmental conditions. Examining Eq. (1), (2), and (3), we can find that the evaporation temperature is the critical factor for evaporation efficiency, because other factors influence the evaporation efficiency by eventually changing the evaporation temperature. Different solar fluxes would just lead to different evaporation temperature. Thus, it is vital to know the relationship between the evaporation efficiency and evaporation temperature.

Interfacial evaporation occurs at the evaporation surface, at which the heat transfer process determines the evaporation efficiency[63]. For the evaporation surface, the evaporation and convection processes are two dominant heat transfer processes and other heat transfer processes could be neglected[63], which could also be verified by our heat transfer model. In this regard, the evaporation efficiency is basically determined by $\eta_e \approx q_{evap}/(q_{evap} + q_{conv})$, which could be regarded as the upper limit of the evaporation efficiency. The natural convection heat loss is a parasitic heat loss associated with evaporation process. The evaporation energy $q_{evap}$ and convection energy $q_{conv}$ are both increasing over temperature, as the inset of Fig. 4b shows. Therefore, for interfacial evaporation, the evaporation efficiency is actually a competition of evaporation energy and convection loss under different evaporation temperature. We could quantify the competition using a ratio of evaporation energy to convection energy ($q_{evap}/q_{conv}$). When the ratio increases, the evaporation efficiency would increase. Thus, it is essential to know how the ratio change when changing the evaporation temperature.

The evaporation and convection energy could be obtained by calculating Eq. (1), (2), and (3). Figure 4b shows the ratio of evaporation energy to convection energy as a function of evaporation temperature for different environmental humidity. When $\varphi = 20\%$ and 60%, when the evaporation temperature increases, the ratio decreases first and then increases. When the environmental humidity increases, the turning points of the lines decreases. When the environmental humidity reached saturation ($\varphi = 100\%$), the ratio monotonously increases as the evaporation temperature increases. These results are consistent with the relationships between evaporation efficiency and solar flux as discussed above, which confirm that the evaporation temperature is actually the underlying mechanisms for evaporation process. To further understand our results, the evaporation process is interpreted in details.

Evaporation is a typical mass transfer process driven by the concentration (or humidity) gradient of the water vapor. When the evaporation temperature is low (nearly identical to the environmental temperature) and the environmental relative humidity is not saturated, there is an inherent humidity gradient between the evaporation surface and the environment, as Eq. (2) shows. In this period, the convection mass transfer is mostly driven by this inherent concentration gradient of the water vapor and the convection heat transfer is negligible, because the temperature difference between the evaporation surface and the environment is nearly equal to 0. Thus, the ratio of evaporation energy to convection energy is extremely high and reaches infinity when the temperature difference is 0. When the evaporation temperature is higher than the turning point, the original concentration gradient is negligible and the concentration gradient of the water vapor is mostly caused by the temperature gradient between the evaporation surface and the environment. Meanwhile, the evaporation energy increases faster than the convection energy when the temperature increases, as the inset of Fig. 4b shows. Thus, increasing the evaporation temperature could increase the ratio when the evaporation temperature is higher than the turning point. However, when the environmental relative humidity is saturated, the original concentration gradient is 0 and the convection mass transfer is only driven by the temperature gradient. Thus, the ratio monotonously increases as the evaporation temperature increases.

From above analysis, the ratio of evaporation energy to convection energy could successfully explain the relationship between the evaporation efficiency and evaporation temperature. Our results could also reveal the underlying mechanisms that changing solar flux could have different effects on the evaporation efficiency, which could help us better understand the conflicting viewpoints about the evaporation temperature. In unsaturated environment, decreased solar flux or evaporation temperature may have higher evaporation efficiency, while only lead to lower evaporation efficiency in saturated environment. Thus, we believe some evaporation structures in previous studies[32,34,43] that can achieve higher evaporation efficiency by increasing evaporation surface area cannot be utilized to harvest fresh water, and would even decrease the evaporation efficiency in actual applications, such as the solar still. It is left for later discussion in section 3.4.

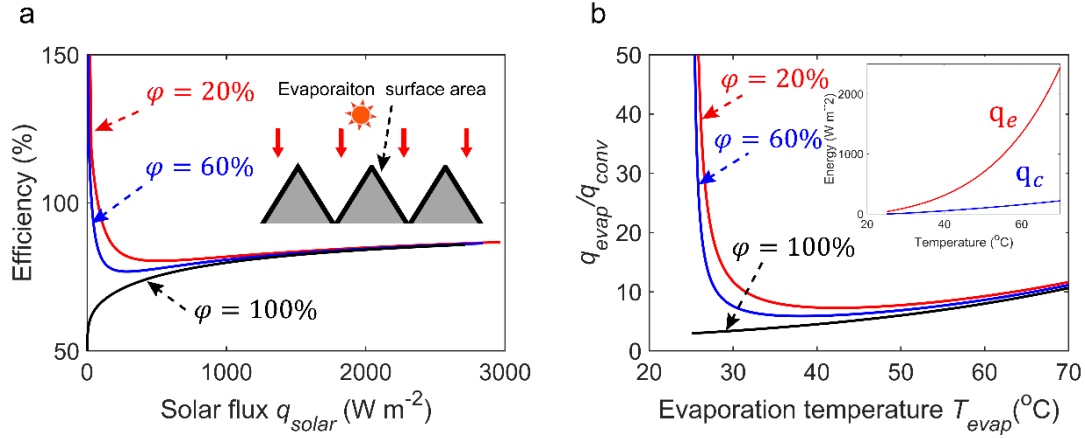

**Figure 4. Evaporation performance under different solar flux and evaporation temperature.** (a) Evaporation efficiency under different solar flux. The insert shows the evaporation structure with increased evaporation surface area. (b) Ratio of evaporation energy to convection energy on the evaporation interface under different evaporation temperature. The insert shows the evaporation energy and convection energy under different temperature.

### 3.3 Reasonable evaluation of evaporation efficiency

In order to fairly evaluate the evaporation efficiencies under different environmental conditions, it is essential to figure out how different environmental conditions influence the evaporation efficiency. The environmental temperature and humidity are the two major environmental factors influencing the evaporation performance, as inferred in Eq. (1), (2), and (3). To compare the evaporation performances of different structures reported by different groups, most previous studies eliminated the effect of environmental conditions by generally subtracting the dark evaporation (i.e., the evaporation without solar) from the total evaporation under solar to obtain the "pure solar-induced evaporation"[13,45]. However, some studies indicated that the dark evaporation should not be a part of the evaporation under solar[34,46].

Our theoretical model provides an effective way to quantify the relationship among the dark evaporation, evaporation under solar, and "pure solar-induced evaporation". The dark evaporation could be obtained by assuming the solar density $q_{solar} = 0$ in Eq. (1) and the evaporation under solar could be obtained by assuming the solar density $q_{solar} = 1$ kW/m$^2$.

Figure 5a shows the impact of environmental temperature to the evaporation rate, assuming the environmental relative humidity 45%. The evaporation rate under solar is found to increase when the environmental temperature increases. However, the "pure solar-induced evaporation" rate decreases when the environmental temperature increases. The results indicate that the "pure solar-induced evaporation" rates are not identical under different environmental temperatures, which means we could not simply eliminate the environment influence by presenting the "pure solar-induced evaporation". Figure 5b shows the impact of environmental relative humidity to the evaporation rate, assuming the environmental temperature 25 ℃. The "pure solar-induced evaporation" rate increases when the environmental relative humidity increases. The results also indicate that the "pure solar-induced evaporation" rates are not identical under different environmental relative humidity and we could not simply eliminate the environment influence by presenting the "pure solar-induced evaporation". It should be mentioned that the evaporation rate under solar decreases very slightly (almost identical) when the environmental relative humidity increase, which means the environmental humidity may not have an impact on the evaporation under solar as much as the environmental temperature.

From above analysis, we can conclude that we could not eliminate the influence of the environmental temperature and humidity by subtracting the dark evaporation rate from the evaporation rate under solar. It seems difficult and sometimes confusing to evaluate and compare the evaporation efficiency of different structures. Thus, it is necessary to have a unified standard of indoor testing conditions[46], to fairly evaluate the evaporation performance of different structures.

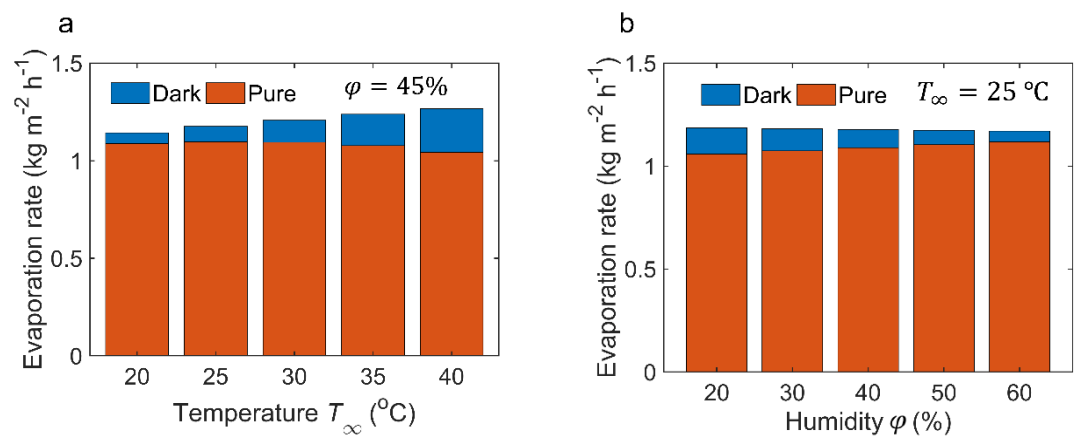

**Figure 5. Effect of environmental temperature and relative humidity to the evaporation rate.** (a) Effect of environmental temperature to the evaporation rate. (b) Effect of environmental humidity to the evaporation rate.

## 3.4 The performance of the interfacial evaporation in actual closed system

From the above analysis, we have discussed the energy efficiency of the interfacial evaporation structure in open space. However, when the interfacial evaporation structure is placed in the enclosed space, such as solar still, the reported system efficiency is only about 30~ 40%[29,31,40-42,47], which means that enclosed system may reduce the evaporation efficiency and other parts of the enclosed system may cause energy loss. In addition, in order to improve the efficiency of water production, it is highly desirable to investigate whether the factors, such as evaporation surface area, discussed above on the evaporation efficiency play the same roles in the solar still. Thus, we further establish a theoretical model of the solar still to analyze the heat transfer in the solar still with the interfacial evaporation, as shown in Fig. 6a. In this model, the solar still is composed of interfacial evaporation structure, which is identical to that is described in previous section on open space evaporation, glass cover and thermal insulation box. Although this model is developed for a solar still, the many conclusions below should be generally applicable to other situations, as long as the system is an enclosed structure.

For the evaporation layer, the energy balance equation Eq. (1) of the interfacial evaporation in open space is followed in enclosed space. It is worth noting that for the enclosed space, part of the solar radiation energy absorbed by the evaporation layer is transmitted to the glass cover through evaporation, convection and radiation, so the heat transfer process of the evaporation layer is affected by the glass cover. In this case, the radiation loss between evaporation layer and glass cover could not be negligible. Also, the convective heat transfer between the glass cover and the evaporation layer needs to consider the influence of limited space on natural convective heat transfer. For the glass cover, the energy balance equation can be expressed as

$$q_{evap} + q_{conv} + q_{rad} - q_{air,in} = q_{conv,g-a} + q_{rad,g-a} + \rho_g c_{p,g} L_g \frac{dT_g}{dt}, \quad (4)$$

where $q_{air,in}$ is the heat loss through the side wall by internal wet air, $q_{conv,g-a}$ is the convection heat transfer flux from glass cover to ambient air. $q_{rad,g-a}$ is the radiation heat transfer flux from evaporation layer to the sky, $c_{p,g}$ is the specific heat of the glass cover, $\rho_g$ is the density of the glass cover, $L_g$ is the thickness of glass cover, $T_g$ is the glass cover temperature.

The temperatures of evaporation layer and the glass cover are the two unknown variables in the energy balance equations of the interfacial evaporation layer and the glass cover. Solving these two equations simultaneously results in the temperatures of the evaporation layer and glass cover as a function of time. The details of theoretical equations and parameters used in solving equations are shown in ESI S2.

To validate the theoretical model, we build a solar still with the same structure and parameters as the model and conducted the solar desalination experiments in Hangzhou, China. The solar radiation intensity, evaporation layer temperature, glass plate temperature and saline temperature of the solar still were recorded from April 10, 2020 to April 11, 2020. The recorded solar intensity and environmental data have been used as inputs of our model. The calculated temperature of the absorber, glass cover and saline, along with the experimental temperature data as a comparison, are shown in Fig. 6b. There is a small time lag between the theoretical results and the experimental data, probably due to the occlusion of sunlight in the afternoon. The overall trend of the temperatures of the different surface are consistent with the experimental results in one day.

According to above analysis on the evaporation structure in Sec. 3.1, it is known that the interfacial evaporation can achieve higher evaporation rate compared with volumetric evaporation. When the water height is 10 cm, the difference of the efficiency for interfacial evaporation and volumetric evaporation can reach 40%. However, in the actual solar still, which is an enclosed space, whether the interfacial evaporation maintains its advantages over the volumetric evaporation needs to be verified. For the volumetric evaporation, the energy balance equation is similar to that of interfacial evaporation, only the heat conduction term is modified. The details of the equations are shown in ESI S2. With the theoretical model, the amount of fresh water productivity in one day is compared between interfacial evaporation and volumetric evaporation in the solar still with same sunlight input and water height, and the results are shown in Fig. 6c. The water productivity of interfacial heating in one day increases very soon and could reach the higher water productivity, about 4.1 kg/(m$^2$·d), while the water productivity of volumetric heating in one day is about 3.1 kg/(m$^2$·d). The difference of water production efficiency between interfacial and volumetric heating in a solar still is smaller (only about 25%) than that in open space (40%). As the relative humidity in the solar still is 100% in the steady state, the vapor pressure difference between the evaporation surface and the glass plate decreases, and the evaporation rate decreases accordingly. Therefore, the advantage of interfacial evaporation is not that significant over the bulk evaporation in the long-term operation of the ordinary solar still.

Also, as discussed in Sec. 3.2, increasing evaporation surface area could increase the evaporation efficiency in the evaporation structure. When the evaporation structure is placed in the solar still, whether changing the evaporation surface area can effectively improve the system efficiency still needs further investigation. According to the models of the evaporation structure and the solar still, we calculate the fresh water production of the two structures with different evaporations surface areas. The red curves of the Fig.6d shows that increasing the evaporation surface area can effectively improve the evaporation efficiency of the evaporation structure. When the evaporation surface area is 0.25 $m^2$, 1 $m^2$ and 3 $m^2$ under the same illumination area, 0.25 $m^2$, the water evaporation of a day is 5.5 $kg/m^2$, 6.2 $kg/m^2$ and 10.3 $kg/m^2$, respectively. However, in the still, the water productivities corresponding to different evaporation surface areas are all about 4.3 $kg/m^2$ in a whole day. Hence, increasing the evaporation surface area seems to have little effect on the system efficiency in an enclosed solar still, even the water productivity has decreased slightly. This is because increasing the evaporation surface area will reduce the solar flux, and the evaporation rate will decrease with the decrease of solar flux as shown in Sec. 3.2. when the relative humidity is nearly 100%. Therefore, for the solar still, we cannot increase the rate of water production by increasing the evaporation surface area.

In order to understand the system efficiency, the energy of the interfacial evaporation structure in the solar still are analyzed, as Fig. 6e shows. For the evaporation, the heat losses are convection loss, radiation loss, conduction loss and optical loss, respectively. The optical loss caused by the light transmittance of the glass plate and the heat conduction loss along the side wall are the main causes of the energy loss, about 28%. Therefore, ensuring the maximum energy input and minimum heat loss is still an important measure to improve the fresh water productivity of the solar still when designing the solar still. Meanwhile, it is necessary to design the solar still which is most suitable for interfacial evaporation to achieve the best water production performance. For example, in view of the characteristics of simple structure and high evaporation efficiency of interfacial evaporation, the multi-stage solar still can be applied to improve the fresh water production[37,38].

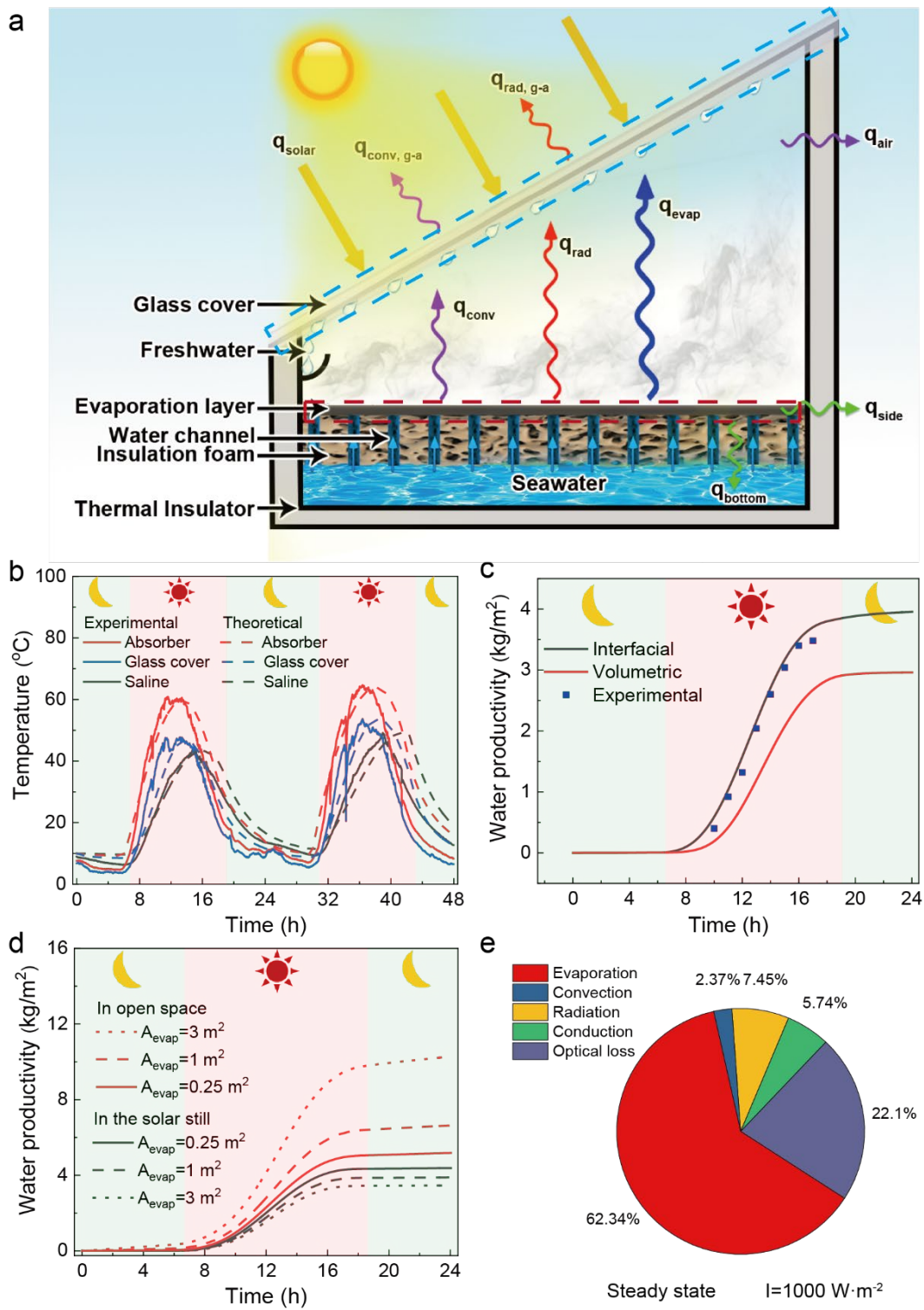

**Figure 6. Performance of interfacial evaporation in solar still.** (a) schematic diagram of the solar still model; (b) Comparison between theoretical calculation and experimental measurement; (c) Water productivity over time under one-day variable sunlight between interfacial heating and the volumetric heating. (d) Water productivity via different evaporation surface areas in the evaporation structure and solar still. (e) Energy distribution in the solar still.

## Conclusions

In conclusion, we have explored the fundamental mechanisms of energy transport in interfacial solar evaporation to interpret different experimental results and conflicting viewpoints.

First, we have identified the advantages of interfacial heating over the traditional volumetric and bottom heating. We show that when the system is well insulated, interfacial heating does not have higher efficiency than the traditional volumetric and bottom heating. The major advantage is that it has much smaller transient heating time (generate steam more quickly at the initial stage of heating) and the relatively less requirement for external thermal insulation.

Second, we have revealed that the evaporation temperature is the determining factor for interfacial solar evaporation efficiency. When the temperature is low, the evaporation is driven by the inherent humidity gradient from the surface to environment. This is the reason why in previous studies, surprisingly high efficiency (more than 100%) can be observed when the surface areas are enlarged. However, this is not more than the well-known evaporative cooling phenomena in mass transfer. Such an enhancement cannot be observed when the evaporation structure is placed in a condition with saturated water vapor (100% relative humidity). On the other hand, when the temperature is high, the evaporation increases more significantly than the convection heat loss, thus leading to a slightly higher efficiency. This indicates that concentrating solar flux can help to boost the efficiency.

Third, we quantitatively illustrate the environmental temperature and humidity to the interfacial evaporation efficiency, which could not be eliminated by subtracting the dark evaporation rate from evaporation rate under solar. To fairly evaluate the efficiency, it is still necessary to have a unified testing condition. Otherwise the efficiencies in different testing conditions cannot be directly compared.

Fourth, although the evaporation rate of interfacial evaporation is very high, the fresh water productivity is still not much higher than a well-designed traditional solar still with bottom heating. Interfacial solar evaporation in an open system and an enclosed system are quite different. Due to the influence of the vapor pressure in the limited space, increasing the evaporation surface area can have negative effect in the solar still.

Our results provide fundamental insights into interfacial solar evaporation and desalination from the energy point of view. Low temperature water vapor is abundant in environment and is basically not useful from the thermodynamics point of view. We suggest that future studies should not be just focusing on the evaporation process itself, but should be more focusing on the full steam generation and water condensation cycle. Based on the energy efficiency analysis, we believe there is still room for further improvement of fresh water productivity. Also, novel strategies to fully utilize the reported high evaporation efficiency of interfacial evaporation are still highly desirable.

## Acknowledgement

We gratefully acknowledge funding support from the Zhejiang Energy R&D Institute and National Natural Science Foundation of China (NO. 51676121).

Supplemental Information for

# The Energy Efficiency of Interfacial Solar Desalination: Insights from Detailed Theoretical Analysis

Xiao Luo[1,5], Jincheng Shi[2,5], Changying Zhao[1], Zhouyang Luo[2,3,4,*], Xiaokun Gu[1,*], Hua Bao[2,*]

[1] Institute of Engineering Thermophysics, School of Mechanical Engineering, Shanghai Jiao Tong University, Shanghai 200240, China

[2] University of Michigan-Shanghai Jiao Tong University Joint Institute, Shanghai Jiao Tong University, Shanghai 200240, China

[3] Key Laboratory of Solar Energy Utilization & Energy Saving Technology of Zhejiang Province, Hangzhou 311121, China

[4] Zhejiang Energy R&D Institute Co., Ltd, Hangzhou 311121, China

[5] These authors contributed equally: X. Luo, J. Shi.

∗ Corresponding authors.

E-mail address: Zhouyang_Luo@outlook.com (Z. Luo); xiaokun.gu@sjtu.edu.cn (X. Gu); hua.bao@sjtu.edu.cn (H. Bao).

## S1 Thermal analysis model for evaporation structure

### (1) Natural convection heat transfer coefficient

To calculate the heat flux due to natural convection (Eq. (2) in the main text,), one needs to determine the natural convection heat transfer coefficient between water surface and environment, $h_{conv}$. $h_{conv}$ can be calculated by Eq. (S1)[1],

$$h_{conv} = Nu \cdot \frac{k_f}{L}, \tag{S1}$$

where $L$ is the characteristic length of the evaporation surface (defined as $A_s/P$, where $A_s$ is the evaporation surface area, $P$ is the evaporation surface perimeter), $k_f$ is the thermal conductivity of the humid air, and $Nu$ is the Nusselt number[1], which can be expressed as,

$$Nu = C(Gr \cdot Pr)^n, \tag{S2}$$

where $Gr$ is the special Grashof number in Dunkle's model[2], $Pr$ is the Prandtl number, C and n is the constants determined by the experimental conditions. According to the experimental steady-state evaporation results, we assume C=0.21 and n=1/4.

**(2) Side & bottom conduction heat losses**

Figure S1 shows the Sectional view of the evaporation structure for interfacial heating (Figure S1 a) and volumetric heating (Figure S1 b). By using the series thermal resistance model[1], for interfacial and volumetric heating, the side conduction heat loss per unit evaporation area can be both expressed as,

$$q_{side} = \frac{A_{side}}{A_{evap}} \cdot \frac{(T_{evap} - T_\infty)}{\frac{l_{ins}}{k_{ins}} + \frac{1}{h_{ins}}}, \qquad (S3)$$

where $A_{side}$ is the total side surface area of the evaporation layer, $A_{side} = 4 * L_e * L_w$, where $L_e$ if the side length of the square evaporation surface and $L_w$ is the evaporation water layer thickness, $A_{evap}$ is the square evaporation surface area, $A_{evap} = L_e * L_e$, $T_{evap}$ is the evaporation temperature, $T_\infty$ is the environmental temperature, $l_{ins}$ is the side thermal insulation thickness, $k_{ins}$ is the insulation thermal conductivity, $h_{ins}$ is the convective coefficient between the insulation surface and environment.

The bottom conduction heat losses are different for interfacial heating and volumetric heating, as Fig. S2 shows. By using series thermal resistance model, for volumetric heating, the bottom conduction heat loss per unit evaporation area can be expressed as,

$$q_{loss} = \frac{(T_e - T_\infty)}{\frac{l_{ins}}{k_{ins}} + \frac{1}{h_{ins}}}. \qquad (S4)$$

By using parallel thermal resistance model, for interfacial heating, the bottom conduction heat loss per unit evaporation area can be expressed as,

$$q_{loss} = \frac{(T_e - T_\infty)}{\frac{l_{foam}}{\emptyset \cdot k_w + (1 - \emptyset) \cdot k_{foam}}}, \qquad (S5)$$

where $l_{foam}$ is the thickness of the insulation foam beneath the evaporation water, $k_{foam}$ is the thermal conductivity of insulation foam, $\emptyset$ is the ratio of water channel area to evaporation surface area, and $k_w$ is the water thermal conductivity. The parameters used in our model is listed in Fig. S1.

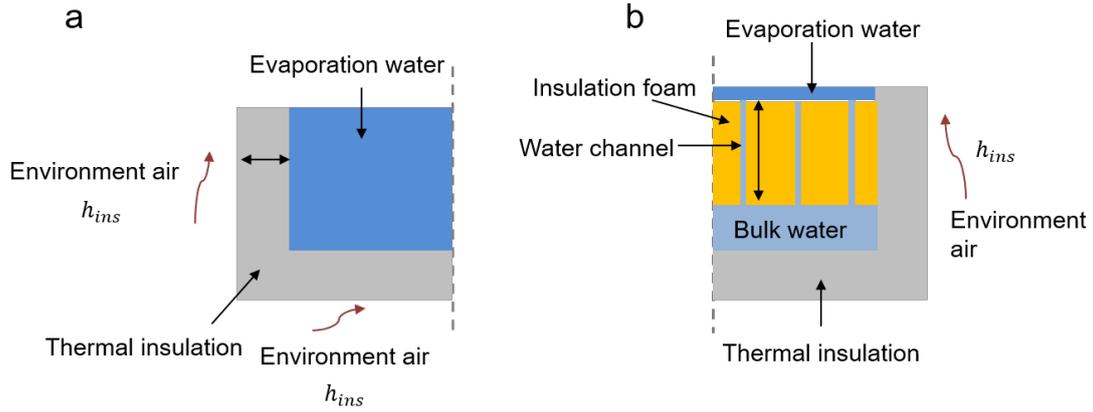

**Figure S1. Sectional view of the evaporation structure. a) Volumetric heating. b) Interfacial heating.**

**Table S1. Parameters' value used in our calculations**

| Side length of the square evaporation surface | $L_e$ | 4 cm |
|---|---|---|
| Absorptivity of the solar absorber | $\alpha$ | 0.97 |
| Thickness of the thermal insulation | $l_{ins}$ | 1 cm |
| Convection coefficient between thermal insulation and environment | $h_{ins}$ | 10 W/(m²·K) |
| Thickness of the insulation foam for interfacial heating | $l_{foam}$ | 2 cm |
| Water channel ratio for interfacial heating | $\emptyset$ | 1% |
| Thermal conductivity of the insulation foam for interfacial heating | $k_{foam}$ | 0.03 W/(m·K) |
| Environmental temperature | $T_\infty$ | 25 °C |
| Environmental relative humidity | $\varphi$ | 0.45 |

## S2 The theoretical model of the solar still

According to the theoretical model, we could write down two heat balance equations for evaporation layer and glass cover in transient state. For the evaporation layer, part of the solar radiation energy absorbed by the evaporation layer is transmitted to the glass cover through evaporation, convection and radiation, part is lost to the side walls and the bulk water, and the remaining part heats up the evaporation layer, the corresponding heat balance equation is expressed as,

$$c_{p,b}m_b\frac{dT_b}{dt} = \alpha_b\tau_g A_b I_0 - h_{conv,b-g}A_b(T_b - T_g) - \varepsilon_{eff}\sigma A_b(T_b^4 - T_g^4) - h_{evap,b-g}A_b(T_b - T_g)$$

$$- A_{s1}\frac{(T_b - T_w)}{\frac{l_{ins}}{k_{ins}} + \frac{1}{h_{ins}}} - A_w\frac{(T_b - T_w)}{\frac{l_{foam}}{\emptyset k_w + (1-\emptyset)k_{foam}}}. \tag{S6}$$

where $c_{p,e}$ is the specific heat of the evaporation surface, $m_b$ is the mass of the evaporation surface, $T_b$ is the evaporation layer temperature, $\alpha_b$ is the solar absorptivity, $\tau_g$ is the transmittance of glass cover, $A_b$ is the evaporation area, $I_0$ is the solar flux, $h_{conv,b-g}$ is the convective heat transfer coefficient between evaporation surface and the glass cover, $T_g$ is the temperature of the glass cover, $\varepsilon_{eff}$ is the effective emittance between the evaporation surface and the lower surface of the glass cover, $\sigma$ the Stefan-Boltzmann constant, $h_{evap,b-g}$ is the evaporation heat transfer coefficient, $A_{s1}$ and $A_w$ are the side and evaporation area, $T_w$ is the bulk water temperature, $l_{iso}$ is the side thermal isolation thickness, $k_{iso}$ is the isolation thermal conductivity, $h_{iso}$ is the convective coefficient between the isolation and environment, $l_{foam}$ is the polystyrene foam thickness, $k_{foam}$ is the thermal conductivity of polystyrene foam, $\emptyset$ is the ratio of water channel area to evaporation surface area, $k_w$ is the water thermal conductivity.

For the volumetric evaporation, the heat balance equation is similar to the model of interfacial evaporation, only the heat conduction term is modified, the equation is expressed as,

$$c_{p,v}m_v\frac{dT_v}{dt} = \alpha_b\tau_g A_b I_0 - h_{conv,b-g}A_b(T_b - T_g) - \varepsilon_{eff}\sigma A_b(T_b^4 - T_g^4) - h_{evap,b-g}A_b(T_b - T_g)$$

$$- A_{s1}\frac{(T_b - T_\infty)}{\frac{l_{ins}}{k_{ins}} + \frac{1}{h_{ins}}} - A_w\frac{(T_b - T_\infty)}{\frac{l_{ins}}{k_{ins}}}. \tag{S7}$$

where $c_{p,v}$ is the specific heat of the bulk water with light absorbing material, $m_v$ is the mass of the bulk water. Other parameters are same to interfacial evaporation.

For the glass cover, the heat transferred from the evaporation surface to the glass plate is equal to the heat transfer between the glass cover plate and the external environment and the temperature rise of the glass cover, the heat balance equation can be expressed as

$$c_{p,g}m_g\frac{dT_g}{dt} = \alpha_g A_g I_0 + h_{conv,w-g}A_b(T_b - T_g) + \varepsilon_{eff}\sigma A_b(T_b^4 - T_g^4) + h_{evap,b-g}A_b(T_b - T_g)$$

$$- \varepsilon_g\sigma A_g(T_g^4 - T_{sky}^4) - h_{conv,g-a}A_g(T_g - T_a) - A_s\frac{(T_{ag} - T_a)}{\frac{\delta_{ins}}{k_{ins}} + \frac{\delta_{iro}}{\lambda_{iro}}} \tag{S8}$$

where $m_g$ is the mass of the glass cover, $c_{p,g}$ is the specific heat of the glass cover, $\alpha_g$ is the solar absorptivity of the glass cover, $\varepsilon_g$ is the emittance of the glass cover, $A_g$ is the glass cover area, $T_{sky}$ is the temperature of the sky, which is assumed to be 3 °C lower than the ambient temperature, $h_{conv,g-a}$ is the convective heat transfer coefficient between upper surface of the glass caver and the ambient environment, $T_a$ is the ambient temperature, $A_s$ is the side area of the solar still, $T_{ag}$ is the average temperature of moist air in solar still, $\delta_1$ is the polystyrene foam thickness, $\lambda_1$ is the thermal conductivity of polystyrene foam, $\delta_2$ is the thickness of iron sheet, $\lambda_2$ is the thermal conductivity of the iron sheet.

For this model, the key to accurately solving the two equations is the determination of the convective heat transfer and evaporation heat transfer between the evaporation layer and the glass cover.

For the convective heat transfer,

$$q_{conv} = h_{conv,b-g}(T_e - T_g), \tag{S9}$$

where $h_{conv,b-g}$ is the convective heat transfer coefficient. In an enclosed space where heat and mass transfer occur simultaneously, the natural convection can be correlated in terms of special Nusselts and Grashof numbers. the correlation between the $h_{conv,b-g}$ and Grashof numbers is[2]

$$h_{conv,b-g} = \frac{Nu \cdot k_f}{\delta} = \frac{C(Gr \cdot Pr)^n \cdot k_f}{\delta}, \tag{S10}$$

where $\delta$ is the plate spacing, $k_f$ is the thermal conductivity of the air.

In this solar still, the characteristic size of the space is 0.3 m, equivalent temperature difference is about 39 °C, then the Grashof number can be calculated about 628650.02. In the horizontal enclosed air space for $3.2 \times 10^5 < Gr < 10^7$, $C$ is 0.075, $n$ is 1/3.

In the enclosed space, the convection heat transfer and evaporation heat transfer occur simultaneously in the evaporation process. Yeh[3] proposed the correlation between convection heat transfer coefficient and evaporation heat transfer coefficient, which had been widely used. Therefore, in this work, we also employ this formula to express the evaporation heat transfer coefficient[4].

$$h_{evap,b-g} = 0.016273 \times \frac{h_{conv,b-g}(p_b - p_g)}{T_b - T_g}, \tag{S11}$$

where $p_b$ is vapor pressure on evaporation surface, $p_g$ is vapor pressure on glass cover.

By solving Eq. (S1) and (S2) the evaporating surface temperature, glass cover temperature as a function of time could be obtained, through the water production rate $\dot{m}$ can be calculated through,

$$\dot{m} = h_{evap,b-g} \frac{T_b - T_g}{h_{fg}}, \tag{S12}$$

where $h_{fg}$ is latent heat of vaporization on the $T_g$.

The parameters used in solving the equation are shown in Table 2.

Table 2. Parameters' value used in our calculations

| For the evaporation layer, | | |
|---|---|---|
| Specific heat of light absorbing materials | $c_{p,b}$ | 4170 J/(kg·K) |
| Density of light absorbing materials | $\rho_b$ | 300 kg/m³ |
| Thickness of light absorbing materials | $L_b$ | 4 cm |
| solar absorptivity | $\alpha_b$ | 0.97 |
| transmittance of glass cover | $\tau_g$ | 0.85 |
| evaporation area | $A_b$ | 0.25 m² |
| the emittance of evaporation layer | $\varepsilon_b$ | 0.95 |
| the emittance of glass cover | $\varepsilon_g$ | 0.92 |
| Stefan-Boltzmann constant | $\sigma$ | 5.67*10⁻⁸ |

| side thermal isolation thickness | $l_{ins}$ | 1 cm |
|---|---|---|
| isolation thermal conductivity | $k_{ins}$ | 0.02 W/(m·K) |
| convective coefficient between the isolation and environment | $h_{ins}$ | 10 W/(m²·K) |
| polystyrene foam thickness | $l_{foam}$ | 2 cm |
| thermal conductivity of polystyrene foam | $k_{foam}$ | 0.03 W/(m·K) |
| ratio of water channel area to evaporation surface area | $\emptyset$ | 1 % |
| For the volumetric evaporation | | |
| Specific heat of volumetric water | $c_{p,v}$ | 4170 J/(kg·K) |
| Density of volumetric water | $\rho_v$ | 1000 kg/m³ |
| For the glass cover, | | |
| specific heat of the glass cover | $c_{p,g}$ | 800 J/(kg·K) |
| solar absorptivity of the glass cover | $\alpha_g$ | 0.1 |
| Glass cover thickness | $L_g$ | 3 mm |
| thickness of iron sheet | $\delta_{iro}$ | 1 cm |
| thermal conductivity of the iron sheet | $k_{iro}$ | 15 W/(m·K) |

# SI Appendix References